\documentclass[a4paper]{jpconf}
\usepackage{graphicx}
\usepackage{amsmath,amssymb}
\usepackage{moreverb}
\usepackage{color}
\newcommand{\be}{\begin{equation}}
\newcommand{\ee}{\end{equation}}
\newcommand{\bea}{\begin{eqnarray}}
\newcommand{\eea}{\end{eqnarray}}

\newcommand{\nn}{\nonumber}
\newcommand{\Den}{{\cal D}}

\newcommand{\SetComplex}{\mathbb{C}}
\newcommand{\Inc}[2]{\left(#1\right)_{#2}}

\begin{document}
\title{One loop integration with hypergeometric series by using recursion relations}

\author{Norihisa Watanabe}

\address{High Energy Accelerator Research Organization (KEK),  Computing Research Center,
1-1, O-ho, Tsukuba, 
Ibaraki 305-0801, Japan}

\ead{norihisa@post.kek.jp}

\author{Toshiaki Kaneko}

\address{High Energy Accelerator Research Organization (KEK), Computing Research Center,
1-1, O-ho, Tsukuba, 
Ibaraki 305-0801, Japan}

\ead{toshiaki.kaneko@kek.jp}

\begin{abstract}
General one-loop integrals
with arbitrary mass and kinematical parameters
in $d$-dimensional space-time  are studied.
By using Bernstein theorem, a recursion relation is obtained which
connects $(n+1)$-point to $n$-point functions.
In solving this recursion relation, we have shown that one-loop integrals
are expressed by a newly defined hypergeometric function, which is
a special case of Aomoto-Gelfand hypergeometric functions.

We have also obtained coefficients of power series expansion around
4-dimensional space-time for two-, three- and four-point functions.
The numerical results are compared  with
\texttt{"LoopTools"} for the case of
two- and three-point functions as examples.
\end{abstract}

\section{Introduction}
For discovery  of the beyond  standard model, we need to know the
precise theoretical prediction of  standard model. 
For Large Hadron Collider at CERN and the international linear collider, at
least next-to-leading order(NLO) electroweak corrections are
necessary.  
% The study of scattering amplitudes for quantum field theory have
% required the computation of Feynman integrals. 
However, it is not easy to calculate Feynman integrations with highly
accuracy even for one-loop level. 
There appear many kinematical parameters including masses,
momenta of particles, and space-time dimension. 
This means that the loop integrals are analytic functions with singularities
on a multiple dimensional complex vector space.
It is difficult to obtain the numerical accuracy in all
kinematical region by a simple numerical approach.
Since the numerical stability of an expression is connected to its
analytic properties, suitable analytic expressions of
one-loop integral are still important.

It is known that any loop integrals are expressed by 
GKZ-hypergeometric functions\cite{GKZ}.
However, theses functions are so general extension
of hypergeometric function that it is not easy to obtain numerical values.
It is desirable to find a subset of GKZ-hypergeometric functions which
corresponds to specific loop integrals to be calculated.

The analytic properties of hypergeometric functions, such as position of
singularities, have been investigated for many hypergeometric functions.
Since these singularities correspond to physical singularities or
large cancellations in numerical calculations,
information about singularities helps us to obtain accurate numerical results.

There are various methods to express
one-loop integrals by hypergeometric
functions\cite{Davy,Tarasov,Duplancic:2000sk,Kurihara:2005at,Kaneko:2011ym}.
In this article, we show a method to obtain analytic expressions of
$n$-point functions with arbitrary kinematical parameters.
Our method is based on Bernstein theorem 
\cite{Bern} (see also \cite{Tkachov:1996wh}).
This theorem implies that
for given polynomial $\Den$ of variables $x=(x_1,...,x_n)$,
there exist differential operator $\mathcal{P}$ and polynomial
$b(s)$ of parameter $s$ such that
\be
\mathcal{P}(\partial, x, s) \Den^{s+1}(x)=b(s) \Den^{s}(x),
\ee
where $\mathcal{P}$ is a polynomial of $x$,  differential operator
$\partial = (\partial_1, ..., \partial_n)$, and parameter $s$.
Applying this theorem to the integrands of one-loop integrals,
we obtain a recursion relation.
In solving this relation, we show that one-loop integrals
are expressed by newly introduced hypergeometric function $G_n$.
This function is found to be one of Aomoto-Gelfand hypergeometric
functions (or general hypergeometric functions on complex
Grassmannians) \cite{Aomoto,Gelfand, Grassman}
which make a subset of GKZ-hypergeometric functions.

Starting from our general expression of one-loop integral,
two- and three-point function are re-expressed 
in a linear combination of Gauss hypergeometric function $F$ and
Appell's function $F_3$\cite{EMO}, respectively.
For the case of scalar integral of four-point function,
we have expanded the analytic expression around 4-dimensional
space-time.
The result is expressed by Lauricella's function $F_D$\cite{EMO} up to
the finite order of space-time dimension $d=4-2\epsilon$.  

We have numerically calculated two- and three-point function as
examples and have compared the results with
\verb"LoopTools"\cite{LoopTools}.

\section{Formulation}\label{SecFormalism}
Let us consider the one-loop $(n+1)$-point function denoted by $I_{n+1}$. After performing integration of $\delta$-function, the Feynman parameter integration of $I_{n+1}$
is written as:
\bea
I_{n+1}(s) &=& \int_{\Delta_n} d^n x\; \Den_n^s\label{integrand}
,\\
\Den_n(x_1, ..., x_n) &
= &
- \frac{1}{2} \sum_{j,k=0}^n 
        q_{jk}^2 x_j x_k
  + \frac{1}{2} \sum_{j, k=0}^n (m_j^2 + m_k^2) x_j x_k,\nn\\
&
 =& \frac{1}{2} \sum_{i,j=1}^n A_{ij} x_i x_j + \sum_{i} B_i x_i + C\nn\\
 &=& \frac{1}{2} (A x, x) + (B, x) + C,
\eea
where 
 $m_i$ are masses of the propagators,
 $q_{jk} \equiv -\sum_{i=j+1}^k p_i$ with external momenta $p_i$,  $x_i$ are Feynman  parameters with $x_0=1-\sum_{j\geq0}x_j$.
Here, $\Delta_n$  refers to $n$-dimensional simplex and $A_{ij}\equiv\partial_{i}\partial_{j}\Den_n$,
$B_{i}\equiv\partial_{i}\Den_n(0)$ and $C\equiv\Den_{n}(0)$.
Parameter $s$ depends on the space-time dimension $d$. 
If $d=4-2\epsilon$ are chosen,
we find $s = 1-n-\epsilon$ for the standard scalar model. 

Let us define  operator ${\cal P}$ by 
\be
{\cal P}\equiv -(s+2)\frac{1}{E_n}+\frac{1}{2E_n}(A^{-1}\partial\Den_n,\partial),
\ee
where $E_n=(A^{-1}B,B)/2-C$. We can find the following relation:
\be
{\cal P} \Den_n^{s+1} =(s+1)\Den_n^s.\label{Bernstein}
\ee
This is the explicit expression of
Bernstein theorem for scalar one-loop integral and
polynomial $b$-function is found to be $s+1$. 
Applying
Eq.({\ref{Bernstein}}) to Eq.({\ref{integrand}}) iteratively with
partial integrations, we obtain
\bea
I_{n+1}(s) &
= &J_{n+1, m}(s) + K_{n+1, m}(s)
,\\
J_{n+1, m}(s) &
= &\frac{1}{2 (s+1) E_n} \sum_{j=0}^{m-1} 
    \frac{\Inc{s+n/2+1}{j}}{\Inc{s+2}{j}}
  \nn\\
  &&\times \int_{\Delta_n} d^n x\; 
     \sum_k \partial_k \left\{(A^{-1} \partial \Den_n)_k \Den_n^{s+1}
                  \left(-\frac{\Den_n}{E_n}\right)^j 
       \right\}
,\\
K_{n+1, m}(s) &
=& \frac{\Inc{s+n/2+1}{m}}{\Inc{s+1}{m}}
  \int_{\Delta_n} d^n x \; \Den_n^s \left(-\frac{\Den_n}{E_n}\right)^{m}\label{homogeneous}
,
\eea
where $\Inc{a}{j} \equiv a (a+1) \cdots (a+j-1)$ is Pochhammer's symbol.

If we choose the appropriate  parameter region, the second term $K_{n+1, m}(s)$ goes to be zero in the limit $m\rightarrow \infty$. The first term is surface integration and can be integrated once easily.  
%After the surface integration, remaining integrations are related to $n$-point functions. Thus, we %obtain
The remaining integrations are expressed by $n$-point functions. 
We obtain a recursion relation between $(n+1)$- and $n$-point
functions:
\bea
I_{n+1}(s)  
= \frac{1}{2} \sum_{j=0}^{\infty} 
    \frac{\Inc{s+n/2+1}{j}}{\Inc{s+1}{j+1}}\left(-\frac{1}{E_n}\right)^{j+1}
    \sum_{k=0}^n h_{\rho(n),k} I_{n,\rho(n;k)}(s+j+1)
,\label{OneIteration}
\eea
where coefficients $h_{\rho(n),k}$ are rational functions of kinematical variables. 
The suffix $\rho(n)=\Delta_n$ and  $\rho(n;k)$ represents  a $(n-1)$-dimensional simplex which
is obtained by eliminating $k$-th vertex and faces attaching to the
vertex from the original $n$-dimensional simplex, which appears as a
part of the boundary of  the original integration domain.
In a similar way, we define $\rho(n;k_1,k_2)\equiv
\rho(\rho(n;k_1),k_2)$.  
Using Eq.(\ref{OneIteration}) repetitively,  
Eq.({\ref{integrand}}) eventually depend  only on the  remaining vertex, in which no integration is left. The final formula of $(n+1)$-point 
function is 
\bea
I_{n+1}(s) &
= &\frac{1}{2^n}
   \sum_{k_{n}=0}^{n} 
   \sum_{k_{n-1}=0}^{n-1}
   \cdots
   \sum_{k_{1}=0}^{1}
   \Den(n;k_n,k_{n-1},...,k_1)^s
     h_{\rho(n),k_n} h_{\rho(n;k_n),k_{n-1}} \cdots
       h_{\rho(n;k_n, ..., k_2),k_1}
   \nn\\
&&
 \times   \sum_{j_{n}=0}^{\infty} 
    \sum_{j_{n-1}=0}^{\infty} 
      \cdots
    \sum_{j_{1}=0}^{\infty} 
      \frac{\Inc{s+n/2+1}{j_{n}}}{\Inc{s+1}{j_{n}+1}}
    \frac{\Inc{s+j_{n}+(n-1)/2+2}{j_{n-1}}}{\Inc{s+j_n+2}{j_{n-1}+1}}
      \cdots
   \nn\\
   &&\qquad \frac{\Inc{s+j_{n}+\cdots+j_{2}+1/2+n}{j_{1}}}{
          \Inc{s+j_{n}+\cdots+j_{2}+n}{j_{1}+1}}\nn
\\ &&\times 
    \left(-\frac{\Den(n;k_n,k_{n-1},...,k_1)}{
         E_{\rho(n)}}\right)^{j_{n}+1} 
    \left(-\frac{\Den(n;k_n,k_{n-1},...,k_1)}{
         E_{\rho(n;k_n)}}\right)^{j_{n-1}+1} 
        \cdots\nn
\\ && \qquad
    \left(-\frac{\Den(n;k_n,k_{n-1},...,k_1)}{
         E_{\rho(n;k_n,...,k_2)}}\right)^{j_{1}+1} 
.\label{HypRepresentation}
\eea
Here, $\Den(n;k_n,k_{n-1},...,k_1)$ is the value at the vertex $k_{n+1}$
which does not appear in the list $(k_n,k_{n-1},\cdots,k_1)$. 
The right-hand side of Eq.(\ref{HypRepresentation}) shows that it is
expressed by a kind of hypergeometric series.
We call this hypergeometric series function $G_n$, which is defined by
Eq.(\ref{defGn}) in the next section.
Using this function, one-loop scalar integral becomes 
\bea
I_{n+1}(s) &
= &\frac{1}{2^n(s+1)_n}
   \sum_{k_{n}=0}^{n} 
   \sum_{k_{n-1}=0}^{n-1}
   \cdots
   \sum_{k_{1}=0}^{1}
  \frac{ \Den(n;k_n,k_{n-1},...,k_1)^{s+n}}
  { E_{\rho(n)} \cdots E_{\rho(n;k_n,...,k_2)}}\nn\\
  &&\times
     h_{\rho(n),k_n} h_{\rho(n;k_n),k_{n-1}} \cdots
       h_{\rho(n;k_n, ..., k_2),k_1}
   \nn\\
   &&\times G_n
   \left(\alpha,\beta;\gamma; \left(-\frac{\Den(n;k_n,k_{n-1},...,k_1)}{
         E_{\rho(n)}}\right),
   \cdots ,  \left(-\frac{\Den(n;k_n,k_{n-1},...,k_1)}{
         E_{\rho(n;k_n,...,k_2)}}\right) 
   \right),
\label{GnRepresentation}
\eea
where 
\bea
\alpha=(\underbrace{1,\cdots,1}_n),\quad\beta=(\underbrace{1/2,\cdots,1/2}_{n-1},s+n/2+1),
\quad\gamma=s+n+1.
\eea

\section{$G_n$-functions}
In this section, we  discuss the properties of $G_n$ function.
This function is defined by:
\bea
G_{n}(\alpha, \beta; \gamma; x) 
=
   \sum_{j_n=0}^\infty
   \sum_{j_{n-1}=0}^\infty
      \cdots
   \sum_{j_{1}=0}^\infty
   \frac{\prod_{i=1}^{n} \Inc{\alpha_{i}}{j_{i}}
     \prod_{k=1}^{n}
      \Inc{\sum_{\ell=k}^n\beta_{\ell}}{\sum_{\ell=k}^{n}j_{\ell}}
   }{
     \Inc{\gamma}{\sum_{i=1}^{n} j_{i}} 
     \prod_{k=1}^{n}
     \Inc{\sum_{\ell=k}^n\beta_{\ell}}{\sum_{\ell=k+1}^{n}j_{\ell}}
     \prod_{i=1}^n j_{i}!
   }
   \prod_{i=1}^{n} x_{i}^{j_{i}}
, \label{defGn}
\eea
where $x = (x_1, ..., x_n) \in \SetComplex^n$ are variables, 
$\alpha=(\alpha_1,...,\alpha_n)$ and $\beta=(\beta_1,...,\beta_n)$ are
complex vectors, and $\gamma$ is a complex parameter.
% related to the space-time dimension.   

Euler type integral representation of $G_n$ is obtained as:
 \bea
 G_{n}(\alpha, \beta; \gamma; x) &
=  & 
   \frac{\Gamma(\gamma)}{
         \prod_{j=1}^n \Gamma(\alpha_j)\Gamma(\gamma-\sum_{k=1}^n \alpha_k)}
   \int_{\Delta_{n}} d^{n} u \;
    \prod_{k=1}^{n} u_k^{\alpha_k-1}\nn
\\ && \qquad\times
    \Bigl(1 - \sum_{j=1}^n u_j \Bigr)^{\gamma-\sum_{k=1}^n \alpha_k -1}
   \prod_{k=1}^{n} \Bigl(1 - \sum_{j=1}^{k} x_j u_j \Bigr)^{-\beta_k}.
\label{IntegrationForm}
 \eea
The integrand is a product of powers of linear factors of integration
variables, while the original integrand of $I_{n+1}$ is a power of
quadratic term.
This representation shows that this function is a member of a class of hypergeometric
functions, which are called Aomoto-Gelfand hypergeometric
functions
%\cite{ Aomoto,Gelfand}
 or hypergeometric functions on complex Grassmannians.
 %\cite{Grassman}
This fact and their analytic properties give us important information
for numerical calculation.  
One can easily show the following formulae:
%In this section, we will discuss the properties of $G_n$-functions.
 %The integration form of  $G_n$ is 
 %\bea
 %G_{n}(\alpha, \beta; \gamma; x) &
%=  & 
 %  \frac{\Gamma(\gamma)}{
%         \prod_{j=1}^n \Gamma(\alpha_j)\Gamma(\gamma-\sum_{k=1}^n \alpha_k)}
 %  \int_{\Delta_{n}} d^{n} u \;
  %  \prod_{k=1}^{n} u_k^{\alpha_k-1}\nn
%\\ && \qquad\times
 %   \Bigl(1 - \sum_{j=1}^n u_j \Bigr)^{\gamma-\sum_{k=1}^n \alpha_k -1}
%   \prod_{k=1}^{n} \Bigl(1 - \sum_{j=1}^{k} x_j u_j \Bigr)^{-\beta_k},\label{IntegrationForm}
% \eea
%where $x = (x_1, ..., x_n) \in \SetComplex^n$ are variables and  $\alpha$, $\beta$ and $\gamma$ are related space-time dimension. 
%The integrand is expressed in  a linear equation although
%the original integrands  are quadratic equation of integration variables.
% So that it gives us important information for numerical calculation.
%We can construct not only integral representation but also some various representation and formulae as follows:
\begin{itemize}
%\item series representation
%\bea
%G_{n}(\alpha, \beta; \gamma; x) 
%=
 %  \sum_{j_n=0}^\infty
%   \sum_{j_{n-1}=0}^\infty
%      \cdots
 %  \sum_{j_{1}=0}^\infty
%   \frac{\prod_{i=1}^{n} \Inc{\alpha_{i}}{j_{i}}
  %   \prod_{k=1}^{n}
 %     \Inc{\sum_{\ell=k}^n\beta_{\ell}}{\sum_{\ell=k}^{n}j_{\ell}}
 %  }{
  %   \Inc{\gamma}{\sum_{i=1}^{n} j_{i}} 
 %    \prod_{k=1}^{n}
  %   \Inc{\sum_{\ell=k}^n\beta_{\ell}}{\sum_{\ell=k+1}^{n}j_{\ell}}
 %    \prod_{i=1}^n j_{i}!
  % }
  % \prod_{i=1}^{n} x_{i}^{j_{i}}
%.
%\eea
\item differentiation (with $k$-th unit vector $e_k$)
\bea
\frac{\partial~}{\partial x_\ell} G_{n}(\alpha, \beta; \gamma; x) = \sum_{k=\ell}^n \frac{\alpha_\ell \beta_k}{\gamma}
  G_{n}(\alpha+e_\ell, \beta+e_k; \gamma+1; x).
\label{differentiation}
\eea
\item recursion relation
\bea
G_{n}(\alpha, \beta; \gamma; x) &
=&
   \sum_{j_n=0}^\infty
   \frac{\Inc{\alpha_n}{j_n}\Inc{\beta_n}{j_n}}{\Inc{\gamma}{j_n} j_n!}
   \;
   x_n^{j_n}
   G_{n-1}(\alpha', \beta'; \gamma'; x')\\
&= &  
   \frac{\Gamma(\gamma)}{\Gamma(\alpha_1) \Gamma(\gamma-\alpha_1)}\nn\\
   &&\hspace{-1cm}
 \times  \int_0^1 d w \;
        w^{\alpha_1-1} (1 - w)^{\gamma-\alpha_1-1}
       (1 - x_1 w)^{-\sum_{j=1}^n \beta_j}
       G_{n-1}(\alpha', \beta'; \gamma'; x')
.
\eea
\end{itemize}
Tensor integrals are obtained by differentiating $I_{n+1}$ in terms of
mass parameters.
Eq.(\ref{differentiation}) shows that tensor integrals are also
expressed by $G_n$.

Based on the above results, the problem of calculating one-loop integral is
converted to one of establishing methods
of expansion around 4 space-time dimension and numerical
evaluations.

We show some samples of scalar case in the next section.

\section{Calculations of $n$-point functions}
In this section, we will discuss how to evaluate and expand $G_n$ functions.
Let us discuss in detail for 2-, 3- and 4-point functions separately. 

\subsection{Two-point function}
\vspace{0.3cm}
Let's first consider the two-point function. This is a good example of the understanding of how to evaluate  $G_n$.
From the formula Eq.(\ref{HypRepresentation}), the scalar two-point function is expressed by $G_1$ functions 
 with $\alpha=1$, $\beta=s+3/2$ and
$\gamma=s+2$ .
%It is notice that tensor integrations are obtained by differentiating scalar integrations by mass parameters. It is easy to find that $G_1$ functions in scaler case can be reduced to Gaussian hypergeometric function as follows:
Function $G_1$ is nothing but
Gaussian hypergeometric function:
\bea
G_1(1,s+3/2;s+2;x)&=&\sum_{j=0}^\infty\frac{ \Inc{1}{j} \Inc{s+3/2}{j}}{ \Inc{s+2}{j}}\frac{x^j}{j!}
=F(1,s+3/2,s+2;x).\label{G1v1}
\eea 
Combining the kinematical factor, we obtain
\bea
I_2(s)&=&\frac{(p^2 + m_1^2 - m_2^2) (m_1^2)^{s+1}}{(s+1) E_1} 
   F\left(1, s+\frac{3}{2}, s+2; -\frac{4 p^2 m_1^2}{E_1}\right) \nonumber
\\ &&
+ \frac{(p^2 + m_2^2 - m_1^2) (m_2^2)^{s+1}}{(s+1) E_1} 
   F\left(1, s+\frac{3}{2}, s+2; -\frac{4 p^2 m_2^2}{E_1}\right),\label{I21}
\eea
where $ E_1\equiv(p^2-(m_1+m_2)^2)(p^2-(m_1-m_2)^2) $. 
The case of $s=-\epsilon$ corresponds to the usual dimensional
regularization $d=4-2\epsilon$.  
%However,  it is difficult to evaluate the value around of arbitrary $\epsilon$, since half integer parameter appear in Eq.(\ref{G1v1}). 
It is more convenient for the expansion when half-integer parameter
is converted to integer. By using identities of $F$,
we obtain
\bea
I_2(s)&=&\frac{\xi_-(m_1^2)^s}{s+1}F\left(1,-s;s+2;\frac{\xi_-}{\xi_+}\right)
+\frac{(1-\xi_-)(m_2^2)^s}{s+1}F\left(1,-s;s+2;\frac{1-\xi_-}{1-\xi_+}\right),
\label{I22}
\eea
where $\xi_\pm=(p^2-m_1^2+m_2^2\pm\sqrt{E_1})/(2p^2)$.
%After the transformation, $G_1$ function can be expanded around arbitrary $\epsilon$ by multiple polylogarithmic functions\cite{Moch:2005uc,Kalmykov:2006hu}.
After the conversion, $G_1$ can be expanded
in arbitrary order of  $\epsilon$ with multiple polylogarithmic functions\cite{Moch:2005uc,Kalmykov:2006hu}.

With the known analytic properties of $F$, 
it is found that two-point function may only be singular when $\xi_\pm=0$, $1$, $\infty$ and $\xi_1=\xi_2$. These cases correspond to massless or on-shell limit.
Let us  investigate the case both of masses are taken massless limit  as an example, where $\xi_-\rightarrow0$ and $\xi_+\rightarrow 1$. The first term in Eq.(\ref{I22}) goes to zero, but second term is not well-defined at this  limit. 
%However, we can transform appropriate representation for taking the massless limit with identities,
However, with using identities of $F$,
we can transform the expression into the well-defined form 
at this limit under the condition ${\rm Re}(s)>0$:
\bea
\lim_{m_1,m_2\rightarrow 0}I_2(s)&=&\lim_{m_1,m_2\rightarrow 0}\frac{(-p^2\xi_+)^s}{s+1}F\left(s+1,-s;s+2;\frac{1}{\xi_+}\right)
=(-p^2)^sB(s+1,s+1),
\eea
where $B$ is beta-function.
%This representation is regular for massless limit under the condition $\SetReal(s)>0$.
This means that we can select appropriate representations in terms of kinematical conditions.

\subsection{Three-point function}
\vspace{0.3cm}
% Corresponding $G_n$ function of three-point function is $G_2$ and
Eq.(\ref{GnRepresentation}) shows that scalar
three-point function is obtained as a linear combination of $G_2$.
It is expressed as
\bea
I_3(s)=\frac{1}{(s+1)(s+2)}\sum_{k_1=1}^2\sum_{k_2=1}^3h_{k_1,k_2}
G_{2}\left(\{1,1\},\{1/2,s+2\};s+3;
    x_{1,(k_1,k_2)},x_{2,(k_1,k_2)}\right).
\label{I31}
\eea
Function $G_2$ is equivalent to Appell's function $F_3$:
\bea
G_2((\alpha_1, \alpha_2), (\beta_1, \beta_2), \gamma; x_1, x_2) 
=  (1 - x_1)^{-\alpha_1}
   F_3\Bigl(\alpha_1, \alpha_2,\gamma-\beta_1-\beta_2, \beta_2, \gamma;
            \frac{x_1}{x_1-1}, x_2 \Bigr).
\eea
When $\alpha_1=\alpha_2=1$,
$G_2$ reduces to Appell's function $F_1$.
It is also convenient when the half-integer parameter is transformed to
integer as same as the case of two-point function.
We apply nontrivial identity:
\bea
&&F_1(\alpha,  \beta,\beta',\gamma;x,y) \nn
\\ 
&&\quad= (1+z)^{\alpha}
   F_D\Bigl(\alpha; \alpha-\gamma+1, 1-\alpha, 2\beta-1, \beta', \beta';
            \gamma;
            z, \frac{z}{1+z}, \frac{2 z}{1+z}, \frac{1}{v_{+}}, 
            \frac{1}{v_{-}} \Bigr),
\label{half2int}
\eea 
where 
$z=(1-\sqrt{1-x})/(1 + \sqrt{1-x})$ ,
 $v_{\pm} 
        = (1+\sqrt{1-x})/(y \mp \sqrt{y(y-x)})$
and $F_D$ is on of Lauricella's functions\cite{EMO}.
For the case $s=-1-\epsilon$, Eq.(\ref{I31})  reduces to
\begin{align*}
&
G_2\rightarrow F_1 \text{(half-integer)}\rightarrow 
  F_D{(\rm integer)}\rightarrow\text{(expansion)}\rightarrow
  F_1\text{(\rm integer)}
\\
& \qquad
 \rightarrow 
 \text{multiple polylogarithmic functions in arbitrary order of } 
 \epsilon.
\end{align*}
 Investigating the limit
$\epsilon\rightarrow 0$, $1/\epsilon$ pole appears from
$1/(s+1)$ of Eq.(\ref{I31}) for both massive and massless cases. 
However, this poles canceled out when all contributions are summed up
for the massive case.
So we can obtain the value of integration in this limit.
\subsection{Four-point function}
\vspace{0.3cm}
After performing $\delta$-function integration, three Feynman parameters
remain on the four-point function.

In this case, $G_3$ appears in Eq.(\ref{GnRepresentation}).
% The corresponding $G_n$ function is $G_3$ defined by
\bea
G_{3}(\alpha, \beta; \gamma; x) 
=&&  \hspace{-0.5cm}
   \frac{\Gamma(\gamma)}{
         \Gamma(\alpha_1) \Gamma(\alpha_2) \Gamma(\alpha_3)
         \Gamma(\gamma-\alpha_1-\alpha_2-\alpha_3)}\nn
\\ && \times
   \int_{\Delta_{3}} d^{3} u \;
    u_1^{\alpha_1-1} u_2^{\alpha_2-1} u_3^{\alpha_3-1}
    (1 - u_1 - u_2 - u_3)^{\gamma-\alpha_1-\alpha_2-\alpha_3-1}\nn
\\ && \times
   (1 - x_1 u_1 )^{-\beta_1}
   (1 - x_1 u_1 - x_2 u_2)^{-\beta_2}
   (1 - x_1 u_1 - x_2 u_2 - x_3 u_3)^{-\beta_3}
.
\eea
For scalar integral case, parameters take the values
$\alpha=\{1,1,1\}$, $\beta=\{1/2,1/2,s+5/2\}$, 
$\gamma=s+4$,  $x=\{x_1,x_2,x_3\}$, and $s=-2-\epsilon$ 
for $d=4-2\epsilon$.
This function  can be written in a linear combination of $F_D$ up to
$\cal{O}(\epsilon)$, which corresponds to finite order of $I_4$,
since
\bea
&&G_3\left(\{1,1,1\},
\left\{\frac{1}{2},\frac{1}{2},\frac{1}{2}-\epsilon\right\};
    2-\epsilon,x_1,x_2,x_3\right)
\nn\\
&&=C_1 F_{1}\left(1,1,\frac{1}{2};2;
      \frac{x_2-x_1}{x_2-1},\frac{x_3-x_1}{x_3-1},\right)
   +C_2 F_D\left(1,2\epsilon,1,1,\frac{1}{2},\frac{1}{2};2; 
      \frac{1-\sqrt{1-x_3}}{2}, \frac{1}{\eta_1}, \frac{1}{\eta_2}, 
      \frac{1}{\eta_3}, \frac{1}{\eta_4}\right)
\nn\\
    &&
    +C_3 F_D\left(1,-\epsilon,1,\frac{1}{2};2-\epsilon; 
       x_1,\frac{x_1-x_2}{1-x_2},
       \frac{x_1-x_3}{1-x_3}\right)+{\cal O}(\epsilon^2),
\eea
where
 the coefficients $C_i$'s and $\eta_i$'s are algebraic functions of $x_1$, $x_2$, and $x_3$.
 The half-integer parameters are converted to integers by using extended 
 identities from Eq.(\ref{half2int}).
 %%%%%%%%%%%%%%%%%%%%%%%%%%%%%%%%%%%%%%%%
 \begin{figure}[b]
  \begin{center}
    \begin{tabular}{c}

      % 1
      \begin{minipage}{0.5\hsize}
        \begin{center}
          \includegraphics[scale=0.4]{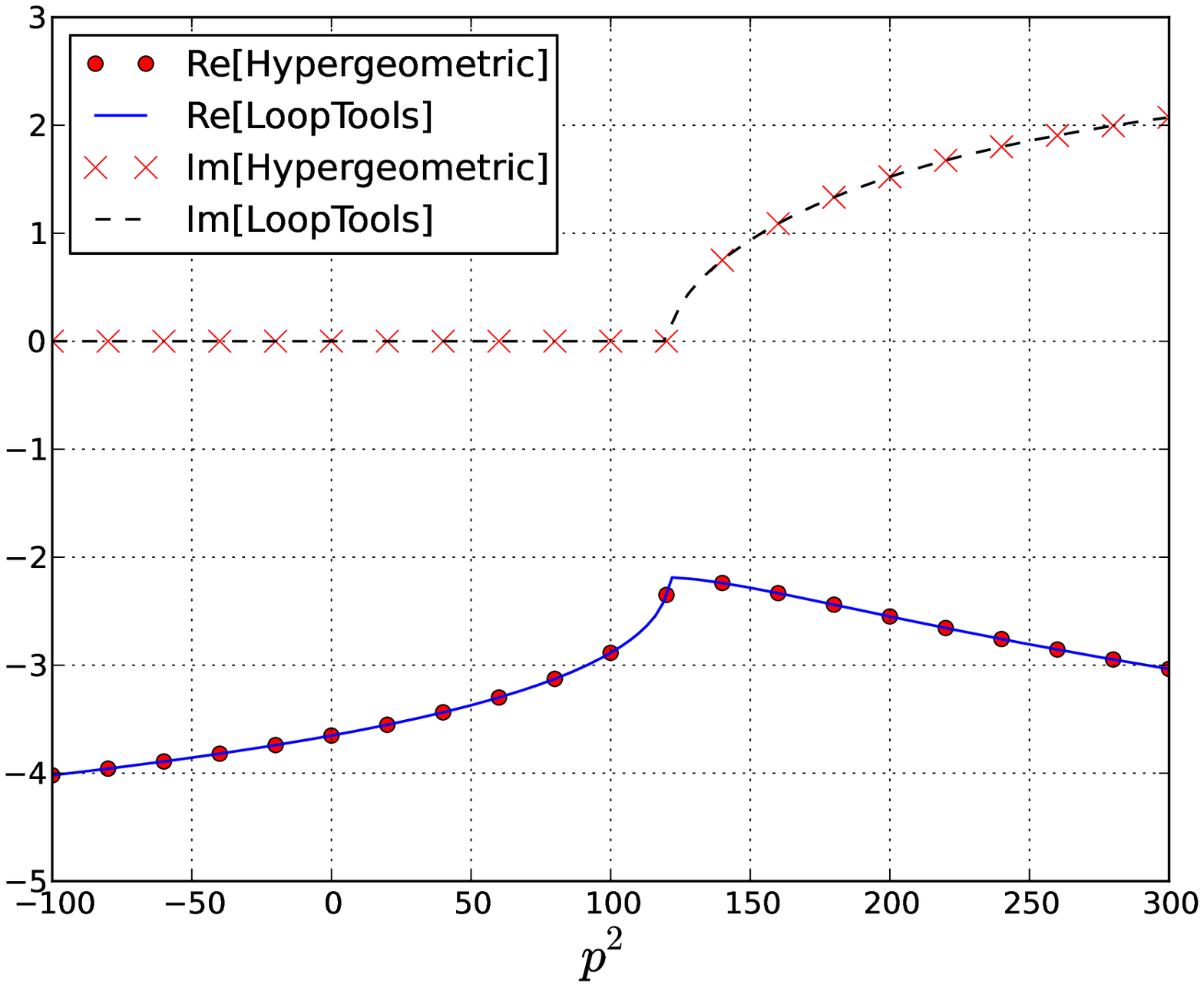}
          \hspace{1.6cm}(a) $m_1^2=10$, $m_2^2=1$
        \end{center}
      \end{minipage}

      % 2
      \begin{minipage}{0.5\hsize}
        \begin{center}
          \includegraphics[scale=0.4]{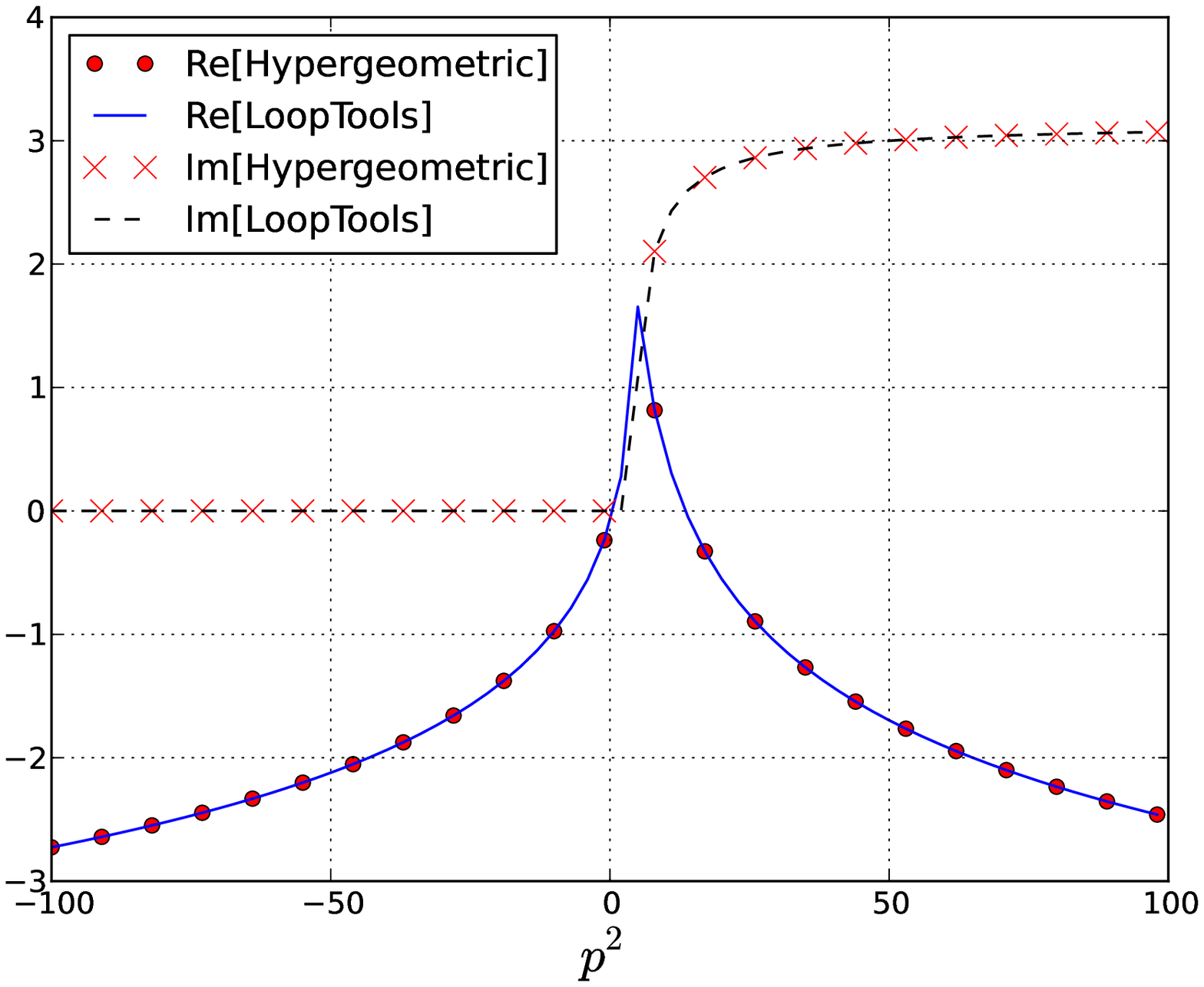}
          \hspace{1.6cm}(b) $m_1^2=1$, $m_2^2=1.1$
        \end{center}
      \end{minipage}
    \end{tabular}
    \caption{
Comparison of numerical results with \texttt{LoopTools}
for two-point functions in two parameter sets (a) and (b). 
Circles and crosses  are the numerical results of real and imaginary part of our
calculation, respectively. 
Solid and dashed  lines are the results of real and
imaginary part which are obtained from  \texttt{LoopTools}, respectively.  
}
    \label{SelfRes}
  \end{center}
\end{figure}
%%%%%%%%%%%%%%%%%%%%%%%%%%%%%%%%%%%%%%%%%%%
 %%%%%%%%%%%%%%%%%%%%%%%%%%%%%%%%%%%%%%%%
 \begin{figure}[htbp]
  \begin{center}
    \begin{tabular}{c}

      % 1
      \begin{minipage}{0.5\hsize}
        \begin{center}
          \includegraphics[scale=0.4]{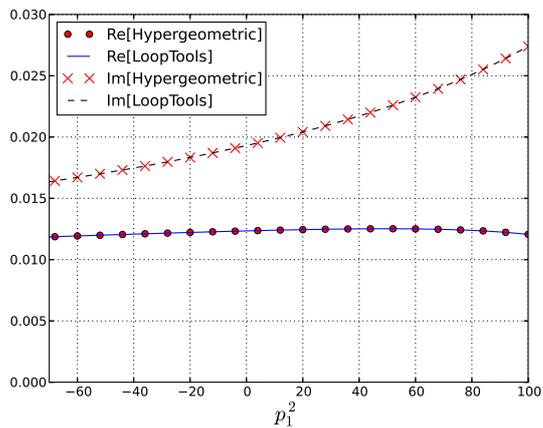}
          \hspace{1.6cm}(a) 
               $p_2^2=-10$, $p_3^2=100$, \\
               $m_1^2=1$, $m_2^2=10$, $m_3^2=100$
        \end{center}
      \end{minipage}

      % 2
      \begin{minipage}{0.5\hsize}
        \begin{center}
          \includegraphics[scale=0.4]{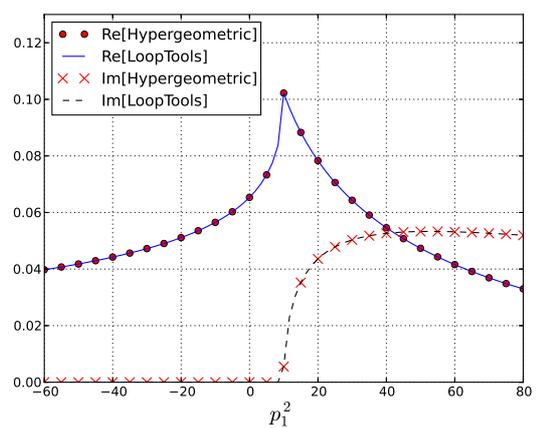}
          \hspace{1.6cm}(b) 
               $p_2^2=-10$, $p_3^2=-100$, \\
               $m_1^2=1$, $m_2^2=2$, $m_3^2=3$
        \end{center}
      \end{minipage}
    \end{tabular}
    \caption{ 
% Compared results of three-point functions with two parameters set (a)
% and (b).   
Comparison of numerical results of three-point functions 
for two parameter sets (a) and (b). 
Circles and crosses  are the numerical results of real and imaginary part of our
calculation, respectively. 
Solid and dashed  lines are the results of real and
imaginary part which are obtained from  \texttt{LoopTools}, respectively.  
}
    \label{VerRes}
  \end{center}
\end{figure}
%%%%%%%%%%%%%%%%%%%%%%%%%%%%%%%%%%%%%%%%%%%

\subsection{Summary numerical calculation}

We have compared the numerical results between our method and
\verb"LoopTools"\cite{LoopTools}. 
% In Figs.(\ref{SelfRes}) and (\ref{VerRes}), 
We show the compared results of two- and  three-point
function in Figs.\ref{SelfRes} and \ref{VerRes}, respectively.  
The results are consistent in satisfactory accuracy.

\section{Conclusion and discussion}

In the discussion of Sec.\ref{SecFormalism}, it is necessary to select appropriate kinematical region in order to make Eq.(\ref{homogeneous})  vanishes at the limit $m\rightarrow \infty$.
However, we can show the following identity of Gauss hypergeometric function:
\be
\Den_n^s=\frac{1}{2(s+1)E_n}\sum_{k}\partial_k
\left[
\left(A^{-1}\partial\Den_n\right)_k
F\left(1,s+\frac{n}{2}+1;s+2;-\frac{\Den_n}{E_n}\right)
\right].
\ee
From this identity, one can derive recursion relation Eq.(9) and confirm it holds in all kinematical region\cite{TKNW}.
% We have shown how to one loop integrations are expressed by
% hypergeometric functions which introduced $G_n$ functions.  

We have shown that general one-loop integral is expressed by $G_n$,
one of  hypergeometric functions on complex Grassmannian.
% to one loop integrations are expressed in
% hypergeometric functions which introduced $G_n$ functions.  
Especially, scalar two- and three-point functions are expressed in terms
of Gaussian and Appell's functions, respectively for any kinematics
variables and space-time dimension. 
Four-point function is expressed in Lauricella's functions up to finite
order for arbitrary kinematical parameters. 
We have also shown the sample numerical calculation in terms of two-,
and three-point functions and results are consistent with
\verb"LoopTools" package.

\section*{Acknowledgment}
The authors wish to thanks to the members of Minami-tateya group for useful discussions.
Especially, we would like to thank to Y. Shimizu and J. Fujimoto for their focus on Bernstein theorem and suggestions.

\end{document}